\newcommand\beq{\begin{equation}}
\newcommand\eeq{\end{equation}}

\documentclass[journal]{IEEEtran}
\usepackage{cite}

\usepackage{graphicx}

\usepackage{amsmath}

\usepackage{array}
\usepackage{dblfloatfix}

\begin{document}
\title{Transformation-Optics-Based Design of a Metamaterial Radome for Extending the Scanning Angle of a Phased Array Antenna}

\author{Massimo~Moccia,
        Giuseppe~Castaldi,
        Giuliana~D'Alterio,
        Maurizio Feo,
		Roberto~Vitiello,  and      
        Vincenzo~Galdi
\thanks{M. Moccia, G. Castaldi and V. Galdi are with the Fields \& Waves Lab, Department of Engineering, University of Sannio, I-82100, Benevento, Italy; e-mail: massimo.moccia@unisannio.it, castaldi@unisannio.it, vgaldi@unisannio.it.}
\thanks{G. D'Alterio, M Feo, and R. Vitiello are with MBDA Italia s.p.a., I-80070 Bacoli (NA), Italy; email: giuliana.dalterio@mbda.it, maurizio.feo@mbda.it, roberto.vitiello@mbda.it}
}

\markboth{}%
{Moccia \MakeLowercase{\textit{et al.}}: Transformation-Optics-Based Design of a Metamaterial Radome for Extending the Scanning Angle of a Phased Array Antenna}


\maketitle

\begin{abstract}
We apply the transformation-optics approach to the design of a metamaterial radome that can extend the scanning angle of a phased-array antenna. For moderate enhancement of the scanning angle, via suitable parameterization and optimization of the coordinate transformation, we obtain a design that admits a technologically viable, robust and potentially broadband implementation in terms of thin-metallic-plate inclusions. Our results, validated via finite-element-based numerical simulations, indicate an alternative route to the design of metamaterial radomes which does not require negative-valued and/or extreme constitutive parameters.
\end{abstract}


\IEEEpeerreviewmaketitle

\section{Introduction}
\IEEEPARstart{T}{he past} two decades have witnessed an exponentially growing interest in electromagnetic (EM) ``metamaterials'' 
\cite{Capolino:2009tp,Cai:2009om,Cui:2009mt}. These are artificial materials, typically consisting of electrically small inclusions in a host medium, engineered so as to attain unconventional EM responses, not necessarily limited by the material properties available in nature. 

From the computational viewpoint, the analysis and design of metamaterials represent quintessential {\em multiscale} problems, characterized by several characteristic sizes spanning orders of magnitude: from the {\em electrically large} size of many operational scenarios of practical interest, through the {\em wavelength-sized} spatial variations of the effective constitutive parameters required in typical designs, up to the {\em deeply sub-wavelength} sizes of inclusions utilized for practical implementations.  
Moreover, after the initial focus on EM, interest has rapidly spread to other disciplines \cite{Kadic:2013fd}, and {\em multiphysics} applications are becoming increasingly relevant \cite{Moccia:2014im,Ma:2014ed,Yang:2015is}.

Metamaterial synthesis has several traits in common with inverse-scattering problems \cite{Colton:2012ia}, and likewise poses some formidable computational challenges. Within the emerging framework of ``metamaterial-by-design'' \cite{Massa:2016ep}, the ``transformation-optics'' (TO) approach \cite{Leonhardt:2006it,Pendry:2006hq} stands out as a systematic strategy to analytically derive the idealized material ``blueprints'' necessary to implement a desired field-manipulation interpreted as a local distortion of the coordinate reference frame. 
Several extensions have also been proposed in order to accommodate effects (e.g., nonlinear, nonreciprocal, 
bianisotropic, magneto-electric, artificial-moving, space-time, nonlocal, non-Hermitian, topological) 
and observables (e.g., resonances, optical forces, density of states) not encompassed by the original formulation (see, e.g., \cite{Bergamin:2008ra,Tretyakov:2008ft,Popa:2011pa,Bergamin:2011pr,Cummer:2011jo,Thompson:2011jo,Castaldi:2012rf,Paul:2012op,Ginis:2012ce,Fernandez:2012to,Castaldi:2013lk,Shi:2015od,Ginis.2015to,Zhang:2016tq,Moccia:2016de}). Moreover, a variety of mechanisms can be exploited in order to {\em reconfigure} the metamaterial response \cite{Oliveri:2015re}.
The reader is also referred to \cite{Werner:2013sp,Kadic:2013fd,Farhat:2016tw} (and references therein) for recent reviews of EM applications as well as extensions to other disciplines. 

In this paper, we apply the TO approach to the design of a metamaterial radome capable of extending the scanning angle of a phased array antenna. It is well know that, in typical phased arrays, the scanning angle is practically limited to $\sim \pm 60^o$ from the broadside direction \cite{Mailloux:2005pa}. In this context, a suitably designed metamaterial radome appears as an attractive alternative to typical mechanical-augmentation systems, in terms of size, weight and complexity.
A first metamaterial-based radome for extending the scanning angle was proposed and successfully realized by Lam {\em et al.} \cite{Lam:2009ni,Lam:2011sp}. Such design is heavily based on brute-force numerical optimization, which results in a nonlinear (and difficult to control) relationship between the input and output angles. Moreover, it relies on negative-index media \cite{Veselago:1968te}, whose metamaterial implementations \cite{Smith:2000cm} are know to be highly dispersive and prone to losses; this inherently curtails bandwidth and efficiency.
In \cite{Sun:2014gm}, Sun {\em et al.} proposed a different, TO-based approach to design two-dimensional (2-D) arbitrarily shaped metamaterial radomes yielding a desired (e.g., linear) relationship between the input and output angles. To overcome the significant complexity of the resulting (anisotropic, inhomogeneous) transformation medium to be implemented, the same authors subsequently put forward a modified approach \cite{Sun:2014kl}, whose implementation relies on the previously introduced ``optical null medium'' \cite{He:2013on}. Though simplified, this implementation still requires material constituents with {\em extreme} parameters: one with very high permittivity and permeability, and the other with near-zero permittivity and permeability.

From the above discussion, it appears that a practical, broadband implementation is not at hand, and there is still considerable room for research. Accordingly, we propose here an alternative TO-based approach that yields constitutive parameters without extreme values (with possibly non-magnetic character) and with moderate anisotropy, which admit
a potentially broadband implementation in terms of inclusions made of thin metallic plates \cite{Chen:2009de}.
More specifically, in Sec. \ref{Sec:Statement}, we describe the problem geometry and outline its conceptual formulation. In Sec. \ref{Sec:Design}, we detail the metamaterial radome design, from the analytically derived, TO-based constitutive blueprints to the actual implementation. In Sec. \ref{Sec:Results}, we illustrate some representative results, and validate our approach via rigorous full-wave numerical simulations. Finally, in Sec. \ref{Sec:Conclusions}, we provide some closing remarks and perspectives.

%
\begin{figure}
\centering
\includegraphics[width=8.5cm]{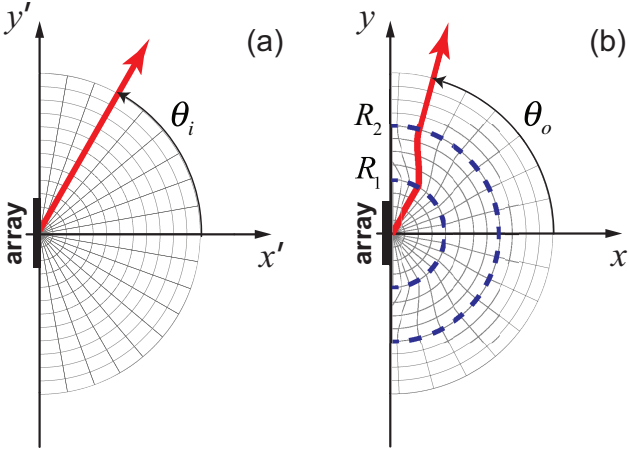}
\caption{Schematic of our approach. (a) In an auxiliary vacuum space, the beam radiated by a phased array (indicated by a thick red arrow) is directed along an angle $\theta_i$. (b) In the transformed space, a coordinate mapping embedded in the annular region $R_1<r<R_2$, $-\pi/2<\theta<\pi/2$ (delimited by the blue-dashed contours) locally distorts the metric so that the same impinging beam is steered toward a different angle $\theta_o>\theta_i$.}
\label{Figure1}
\end{figure}

\section{Problem Geometry and Formulation}
\label{Sec:Statement}
The geometry and general idea underlying our approach are schematized in Fig. \ref{Figure1}. Throughout the paper, we assume an implicit $\exp\left(-i\omega t\right)$ time-harmonic excitation, and a 2-D scenario, with all fields and quantities independent of $z$, and transverse-magnetic (TM) polarization (i.e., $z$-directed magnetic field).
We begin considering an auxiliary vacuum space with coordinates ${\bf r}^\prime\equiv\left(x^\prime,y^\prime,z^\prime\right)$ [and associated cylindrical reference system $\left(r^\prime,\theta^\prime,z^\prime\right)$] where an equivalent aperture field distribution located at the $x^\prime=0$ plane (representative of the phased array antenna) radiates a directive beam
[schematized as a thick arrow in Fig. \ref{Figure1}(a)] along a direction $\theta_i$ with respect to the $x^\prime$ axis. 
Next, we consider a 2-D coordinate transformation to a distorted reference frame ${\bf r}\equiv\left(x,y,z\right)$ [and associated cylindrical reference system $\left(r,\theta,z\right)$]
\beq
{\bf r}^\prime={\bf F}\left({\bf r}\right),
\label{eq:CT}
\eeq
confined to a finite region nearby the aperture. Even though arbitrary shapes are possible, for simplicity, we assume such region to be the annular domain $R_1<r<R_2$, $-\pi/2<\theta<\pi/2$. The above mapping belongs to the general class of ``finite embedded coordinate transformations'' introduced in \cite{Rahm:2008od}. As conceptually illustrated in Fig. \ref{Figure1}(b), it locally distorts the energy flow so that the impinging beam [cf. Fig. \ref{Figure1}(a)] is gradually steered and exits the annular region with a different direction
\beq
\theta_o=\alpha \theta_i,
\label{eq:steering}
\eeq
with $\alpha>1$ denoting a desired enhancement factor.

Following the TO approach \cite{Pendry:2006hq}, and invoking the form-invariance properties of Maxwell's equations, the above effects can be equivalently obtained by filling the annular domain $R_1<r<R_2$, $-\pi/2<\theta<\pi/2$ with an anisotropic, inhomogeneous medium characterized by relative permittivity and permeability tensors
\beq
{\underline {\underline \varepsilon}}\left({\bf r}\right)=
{\underline {\underline \mu}}\left({\bf r}\right)=
\det\left[{\underline {\underline \Lambda}}\left({\bf r}\right)\right]
{\underline {\underline \Lambda}}^{-1}\left({\bf r}\right)
\cdot
{\underline {\underline \Lambda}}^{-T}\left({\bf r}\right),
\label{eq:TO-BP}
\eeq
where $\mbox{det}$, $^{-1}$ and $^{-T}$ denote the determinant, inverse, and inverse transpose, respectively, whereas ${\underline {\underline \Lambda}}\equiv\partial {\bf F}/\partial {\bf r}$ is the Jacobian matrix associated with the coordinate transformation in (\ref{eq:CT}).

In what follows, we address the judicious choice of the coordinate transformation, as well as the practical implementation of the TO-based constitutive ``blueprints'' in (\ref{eq:TO-BP}).

\section{Metamaterial Radome Design}
\label{Sec:Design}

\subsection{Coordinate Transformation}
To implement the manipulation schematized in Fig. \ref{Figure1}, we choose a 2-D coordinate transformation
\begin{IEEEeqnarray}{rCl} 
\label{eq:CTr}
\IEEEyesnumber \IEEEyessubnumber*
r^\prime &=& F_r\left(r\right),\label{eq:CF1}\\
\IEEEyessubnumber
\theta^\prime&=&F_{\theta}\left(r\right)\theta,\\
z^\prime&=&z,
\end{IEEEeqnarray}
with the following boundary conditions
\begin{IEEEeqnarray}{rCl} 
\label{eq:BC}
\IEEEyesnumber \IEEEyessubnumber*
F_r\left(R_1\right)&=&R_1,\label{eq:c1}\\
F_r\left(R_2\right)&=&R_2,\label{eq:c2}\\
F_{\theta}\left(R_1\right)&=&1,\label{eq:c3}\\
F_{\theta}\left(R_2\right)&=&\label{eq:c4}\frac{1}{\alpha}.
\end{IEEEeqnarray}
In particular, (\ref{eq:c1}) and (\ref{eq:c3}) ensure that the transformation reduces to the identity at the interior boundary $r=R_1$, whereas (\ref{eq:c2}) and (\ref{eq:c4}) yield the desired steering condition in (\ref{eq:steering}) at the exterior boundary $r=R_2$. This last condition also implies that, as inherent of finite embedded coordinate transformations \cite{Rahm:2008od}, the above mapping cannot reduce to the identity at the exterior boundary, which in turn indicates that perfect impedance matching {\em cannot} be attained (see also our discussion below).

The coordinate transformation in (\ref{eq:CTr}) [with (\ref{eq:BC})] differs from those in previous approaches \cite{Sun:2014gm,Sun:2014kl} due to the presence of the radial mapping $F_r\left(r\right)$ in (\ref{eq:CF1}), which reduces to the identity at the boundaries $r=R_1$ and $r=R_2$ [cf. (\ref{eq:c1}) and (\ref{eq:c2})], but is generally {\em non-identical} inside the annular domain $R_1<r<R_2$. This introduces additional degrees of freedom, which can be exploited {\em i)} to minimize the anisotropy of the arising transformation medium, and {\em ii)} to control its magnetic character so as, e.g., to render its effective magnetic permeability constant.

%
\begin{figure}
\centering
\includegraphics[width=8.5cm]{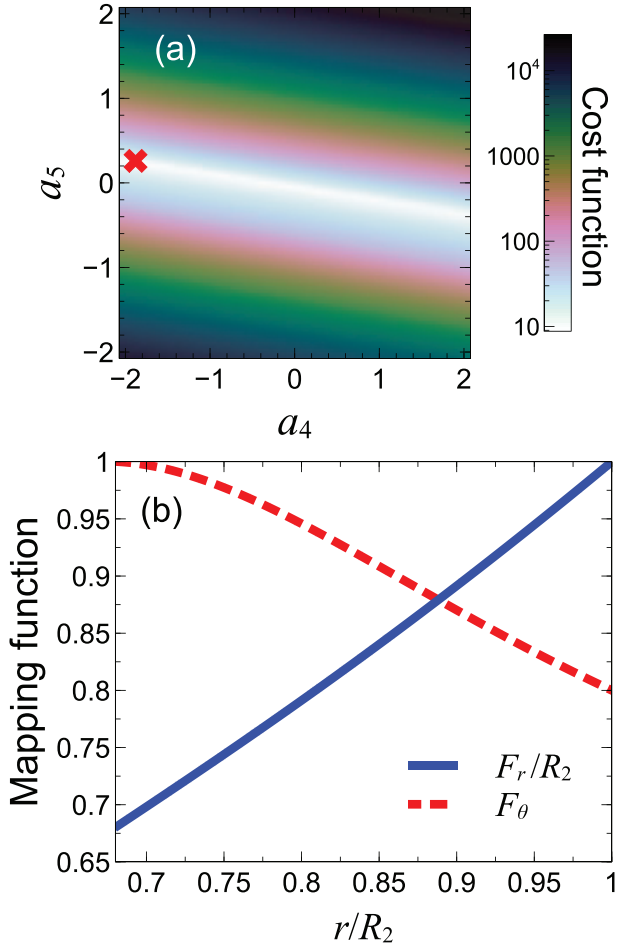}
\caption{Geometry as in Fig. \ref{Figure1}, with $\alpha=1.25$ and $R_2=1.467 R1$, and transformation parameters $\mu_z=0.9$, $a_0=-1.040$, $a_1=5.484$, $a_2=-7.083$, $a_3= 5.238$. (a) Cost function in (\ref{eq:CF}) as a function of the parameters $a_4$ and $a_5$. The red-cross marker indicates the absolute minimum at $a_4=-1.867$ and $a_5=0.267$. (b) Optimized mapping functions $F_r$ (normalized with respect to $R_2$; blue-solid curve) and $F_{\theta}$ (red-dashed curve).}
\label{Figure2}
\end{figure}

\subsection{TO-Based Constitutive Blueprints}
\label{Sec:Blueprints}
By particularizing the general TO-based constitutive blueprints (\ref{eq:TO-BP}) to our chosen coordinate transformations in 
(\ref{eq:CTr}), we obtain for the constitutive tensors (in cylindrical components)
\beq
{\underline {\underline \varepsilon}}=
{\underline {\underline \mu}}=
\begin{bmatrix}
\displaystyle{\frac{F_r F_{\theta} }{r \dot{F_r}}} &
-\displaystyle{\frac{\theta F_r \dot{F_{\theta}}}{\dot{F_r}}}
& 0\\
-\displaystyle{\frac{\theta F_r\dot{F_{\theta}}}{\dot{F_r}}} & 
\displaystyle{\frac{r \left(\dot{F_r}^2+\theta^2 F_r^2\dot{F_{\theta}}^2\right)}{F_r
\dot{F_r} F_{\theta}}} & 
0\\
0 & 0 & \displaystyle{\frac{F_r
\dot{F_r} F_{\theta}}{r}}
\end{bmatrix},
\label{eq:TrasnfM}
\eeq
where the overdot denotes differentiation with respect to the argument and, for notational compactness, we have omitted the explicit dependence on $r$ of the functions $F_r$ and $F_{\theta}$.
The transformation medium described by the constitutive relationships in (\ref{eq:TrasnfM}) is generally inhomogeneous and anisotropic. In view of the assumed TM polarization, we can restrict our attention to the in-plane ($r$, $\theta$) permittivity components and the $z$-component of the permeability. By diagonalizing the tensors in (\ref{eq:TrasnfM}), we obtain a more insightful interpretation in terms of a uniaxially anisotropic medium with real-valued components (in view of the symmetric character of the tensors) and tilted optical axis (depending on the local direction of the eigenvectors). By calculating the eigenvalues of (\ref{eq:TrasnfM}), we obtain that positive values of the permittivity components in this principal reference system can be enforced via the condition
\beq
\frac{\theta^2 r F_r\left(r\right)\dot{F_{\theta}}^2\left(r\right)}{F_{\theta}\left(r\right)\dot{F_r}\left(r\right)} + 
\frac{r\dot{F_r}\left(r\right)}{F_r\left(r\right)F_{\theta}\left(r\right)} 
+ \frac{F_r\left(r\right)F_{\theta}\left(r\right)}{r\dot{F_r}\left(r\right)} > 2,
\label{eq:pos_eps}
\eeq
whereas the condition for positive permeability directly follows from (\ref{eq:TrasnfM}) as
\beq
F_r\left(r\right)
\dot{F_r}\left(r\right) F_{\theta}\left(r\right)>0.
\eeq

As previously mentioned, the coordinate transformation in (\ref{eq:CTr}) [with (\ref{eq:BC})] inherently yields
an impedance mismatch. In what follows, we focus on moderate values of the enhancement factor ($\alpha=1.25$), for which a rough estimate of the associated reflections can be worked out (see Appendix for details), which relates the standing-wave-ratio (SWR) to $\alpha$, viz.,
\beq
\mbox{SWR}\approx \alpha.
\label{eq:SWR}
\eeq
From (\ref{eq:SWR}), we gather that the impedance mismatch should be tolerable for the targeted enhancement factor. However, for significantly larger values of $\alpha$, the above estimate may become inaccurate, and the impedance mismatch no longer tolerable. 

%
\begin{figure}
\centering
\includegraphics[width=8.5cm]{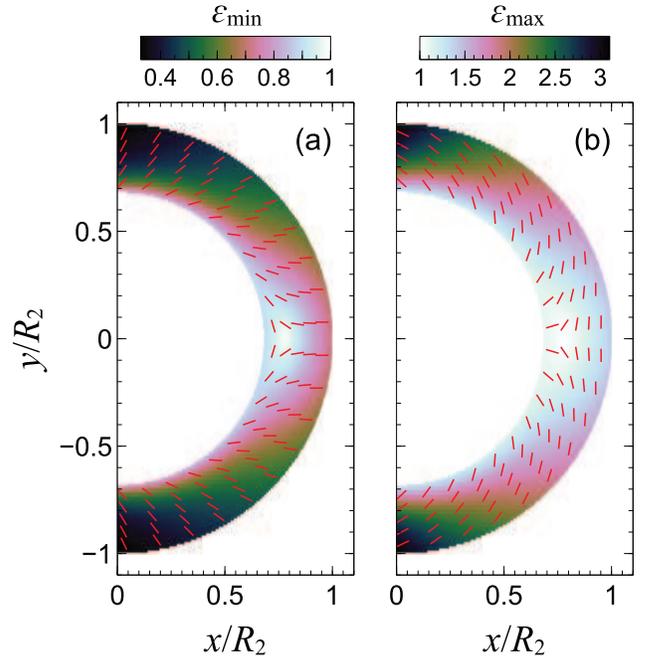}
\caption{Geometry as in Fig. \ref{Figure1}, with $\alpha=1.25$ and $R_2=1.467 R1$, with optimized parameters as in Fig. \ref{Figure2}. 
(a), (b) Spatial distributions of the minimum and maximum relative-permittivity components [in-plane eigenvalue of the tensor in (\ref{eq:TrasnfM})], respectively, in the principal reference system. 
As a reference, the local directions of the principal axes (eigenvectors) are shown as short red segments. Note the inverted colorscales in the two plots.}
\label{Figure3}
\end{figure}

\subsection{Parameterization and Optimization of the Coordinate Transformations}
With a view toward a practical metamaterial implementation in terms of inclusions made of thin metallic plates (along the lines of \cite{Chen:2009de}), it makes sense to enforce upfront some expected characteristics. First, it is know that, for the assumed TM polarization, the above implementation yields metamaterials characterized by almost constant, nearly-one relative permeability \cite{Chen:2009de}. We therefore enforce the condition
\beq
F_{\theta}\left(r\right)=\frac{r \mu_z}{F_r\left(r\right) \dot{F_r}\left(r\right)},
\label{eq:gcond}
\eeq
which, from (\ref{eq:TrasnfM}), yields a constant value $\mu_z$ of the relevant relative-permeability component. The above condition implies that the mapping function $F_r$ and $F_{\theta}$ cannot be chosen independently.
For the mapping function $F_r$, we choose a rather general polynomial parameterization
\beq
F_r\left(r\right)=R_1\sum_{n=0}^{4}a_n \left(\frac{r}{R_1}\right)^n,
\eeq
with $a_n$ denoting dimensionless, real-valued coefficients to be determined. By enforcing the boundary conditions in (\ref{eq:BC}), with the function $F_{\theta}$ calculated from (\ref{eq:gcond}), we obtain a linear system of four equations in the unknown coefficients $a_n$, which we exploit to calculate analytically the lowest-order coefficients ($n=0,...,3$). The remaining two coefficients ($a_4$ and $a_5$), together with the relative permeability $\mu_z$, are in principle free parameters that can be exploited to optimize the transformation. However, anticipating the metamaterial implementation detailed in Sec. \ref{Sec:MTM-synthesis}, we select $\mu_z=0.9$ as a realistically attainable value, and choose the coefficients $a_4$ and $a_5$ by minimizing the anisotropy of the transformation medium. By letting $\varepsilon_{max}$ and $\varepsilon_{min}$ the maximum and minimum permittivity components, respectively, of the relative permittivity tensor in (\ref{eq:TrasnfM}) in the principal reference system (i.e., the in-plane eigenvalues), the optimization problem
can be mathematically formulated as minimizing the cost function
\beq
J\left(a_4,a_5\right)=
\max_{r, \theta}\left|
\frac{\varepsilon_{max}\left(r,\theta\right)}{\varepsilon_{min}\left(r,\theta\right)}
\right|.
\label{eq:CF}
\eeq
Given the low-dimensional character of the optimization space, exhaustive search is a computationally affordable route for the above problem. In our numerical implementation, the annular region of interest $R_1<r<R_2$, $0<\theta<\pi/2$ (due to symmetry) is uniformly sampled on a $21\times 19$ grid. Moreover, we enforce positive values of the permittivity components by discarding candidate solutions that do not satisfy the condition in (\ref{eq:pos_eps}). Figure \ref{Figure2}(a) shows the cost function within the region $a_4, a_5 \in [-2,2]$, for a
configuration with $\alpha=1.25$, $R_2=1.467 R_1$, and the other parameters given in the caption. A rather deep minimum can be observed, which is rather broad along a specific direction in the parameter space. An exhaustive search, with progressive restriction of the search space and refinement of the sampling step (up to 0.001) yields the global minimum indicated with a red-cross marker. Figure \ref{Figure2}(b) shows the resulting optimized mapping functions.

The resulting contitutive parameters are shown in Fig. \ref{Figure3}. More specifically, the two plots show the spatial distributions of the minimum [Fig. \ref{Figure3}(a)] and maximum [Fig. \ref{Figure3}(b)] relative-permittivity components in the principal reference system, whose local axes (eigenvector directions) are shown as short red segments. As it can be observed, parameters are everywhere positive, without extreme values. The anisotropy is moderate, in line with the values that can be attained with practical implementations based on thin-metallic-plate inclusions \cite{Chen:2009de}.

%
\begin{figure}
\centering
\includegraphics[width=8.5cm]{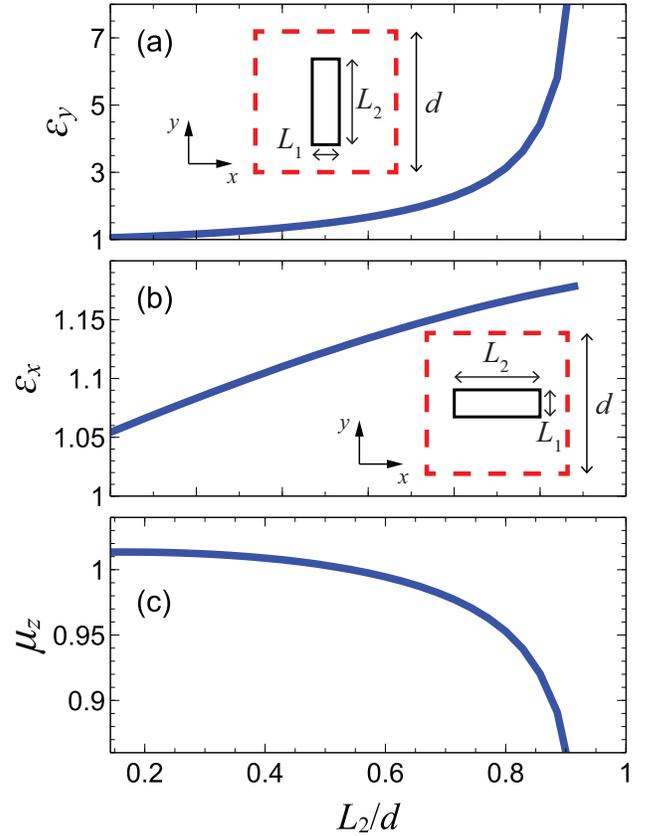}
\caption{Effective constitutive parameters for a metamaterial implementation based on thin-metallic-plate inclusions, at the reference frequency of 14.5 GHz. (a), (b) In-plane relative-permittivity components, as a function of the inclusion sidelength $L_2$, extracted by assuming the unit-cells shown in the insets, with $d=3.5$ mm and $L_1=0.5$ mm. (c) Corresponding out-of-plane relative permeability, averaged over the two unit-cell orientations.}
\label{Figure4}
\end{figure}

%
\begin{figure}
\centering
\includegraphics[width=8.5cm]{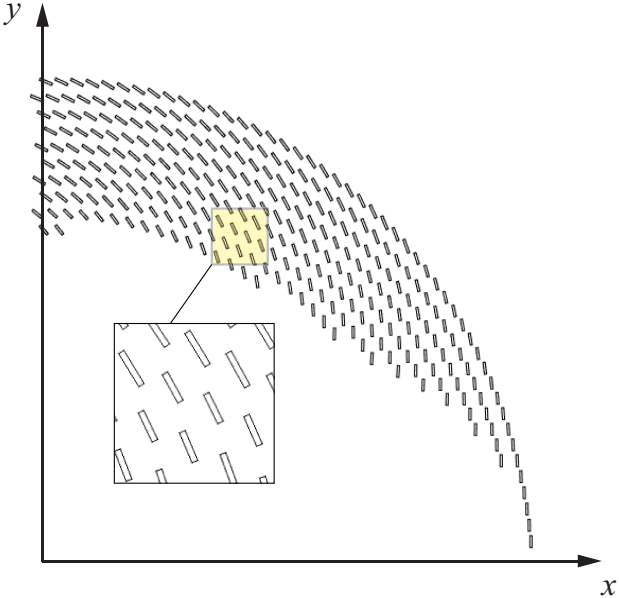}
\caption{Schematic of the metamaterial implementation based on thin-metallic-plate inclusions, approximating the TO-based blueprints in Fig. \ref{Figure3} (with $R_1=75$ mm and $R_2=110$ mm) at the reference frequency of 14.5 GHz. 
The structure is composed of 566 inclusions (270 with $L_2=2.5$ mm, 296 with $L_2=3$ mm; all with $L_1=0.5$ mm).
Due to symmetry, only the upper ($y>0$) half is displayed. The inset shows a magnified view of the yellow-shaded region.}
\label{Figure5}
\end{figure}

\subsection{Metamaterial Synthesis}
\label{Sec:MTM-synthesis}
The TO-based constitutive blueprints in (\ref{eq:TrasnfM}) loosely resemble those considered in \cite{Chen:2009de} in connection with a ``rotation cloak''. It therefore makes sense to follow a similar metamaterial-synthesis approach based on thin-metallic-plate inclusions.

To this aim, as schematically illustrated in Fig. \ref{Figure4}(a) inset, we consider a square unit-cell of sidelength $d$, containing a rectangular metallic inclusion of sidelengths $L_1$ and $L_2$. For this structure, we extract the effective constitutive parameters following the procedure detailed in \cite{Chen:2009de}, which is in turn based on the general approach introduced in \cite{Smith:2002do}. Figure \ref{Figure4} shows the extracted constitutive parameters, assuming a reference frequency of 14.5 GHz, and the rectangular inclusion oriented along the $x$ and $y$ axes. More specifically, to simplify the practical implementation, we fix the dimension $L_1=0.5$ mm, and are therefore left with only one degree of freedom ($L_2$). To extract the constitutive parameters, we perform two full-wave numerical simulations (see Sec. \ref{Sec:Numerical} for details), with the inclusion side of length $L_2$ oriented along the $y$- and $x$- axis [see the insets in Figs. \ref{Figure4}(a) and \ref{Figure4}(b), respectively], and assuming infinite periodicity along the $y$-direction and normally incident, TM-polarized plane-wave illumination. From the computed scattering parameters, we extract the in-plane relative-permittivity components $\varepsilon_y$ and $\varepsilon_x$ [shown in Figs. \ref{Figure4}(a) and \ref{Figure4}(b), respectively, as a function of $L_2$] and the out-of-plane relative-permeability component $\mu_z$. For this latter, as also observed in \cite{Chen:2009de}, the values extracted considering the two orthogonal inclusion orientations are slightly different ($\sim 9$\%, on average);  the values shown in Fig. \ref{Figure4}(c) correspond to the average of these two cases. By comparing the attainable  constitutive parameters in Fig. \ref{Figure4} with the targeted TO-based blueprints in Fig. \ref{Figure3}, the following observations are in order. First, our metamaterial implementation exhibits the anticipated weak magnetic response, with mild dependence on the inclusion size, which justifies our assumption of $\mu_z=0.9$ in the TO-based blueprints. Second, unlike the TO-based blueprints, both relative-permittivity components in our metamaterial implementation are larger than one. Strictly speaking, this implies that an {\em exact} synthesis is not possible. Nevertheless, as also pointed out in \cite{Chen:2009de}, the basic functionality of the device is still preserved if the synthesis is relaxed in such a way that, instead of requiring the exact parameter matching, only the anisotropy ratio ($\varepsilon_{max}/\varepsilon_{min}$) and the optical-axis direction are matched. This relaxed synthesis is compatible with our metamaterial implementation in Fig. \ref{Figure4}, and its inherent imperfection mainly affects the impedance matching.

Accordingly, to synthesize the TO-based blueprints in Fig. \ref{Figure3}, assuming $R_1=75$ mm and $R_2=110$ mm, we discretize the domain of interest with a uniform polar grid with 842 cells of size $\approx 3.5$mm (i.e., $\approx\lambda_0/6$, with $\lambda_0$ denoting the vacuum wavelength at the reference frequency of 14.5 GHz). In each cell, we place a thin-metallic-plate inclusion, with $L_1=0.5$ mm, orientation suitably chosen so as to match the local principal reference system, and the sidelength $L_2$ extracted from Fig. \ref{Figure4} so as to match the local anisotropy ratio. 
This yields a preliminary structure comprising 842 inclusions with sidelength $L_2$ within the range $[0.945~\mbox{mm} - 3.2 ~\mbox{mm}]$. 
To simplify the structure and assess its robustness with respect to fabrication tolerances, we subsequently introduce a five-level quantization of the $L_2$ values (from 1 mm to 3 mm, with step of 0.5 mm), and progressively decimate the smaller inclusions, by monitoring the effects in the simulated EM response (see Sec. \ref{Sec:Numerical} for details).
Figure \ref{Figure5} shows the final outcome of this procedure, in which only 566 inclusions of two different types (270 with $L_2=2.5$ mm, and 296 with $L_2=3$ mm) are retained.

\section{Representative Results}
\label{Sec:Results}

\subsection{Numerical Modeling and Observables}
\label{Sec:Numerical}
Our numerical modeling of the TO-based and synthesized-metamaterial structures rely on the finite-element based commercial software package COMSOL Multiphysics \cite{COMSOL:2015}. 

More specifically, for the metamaterial parameter extraction in Sec. \ref{Sec:MTM-synthesis} (see Fig. \ref{Figure4}), we consider a square unit-cell of size $d=3.5$ mm, with Bloch-type periodicity conditions along the $y$-direction and two port-type terminations placed at a distance of 21 mm along the $x$-direction. For each value of $L_2$, and for the two possible orientations of the unit cell [see Figs. \ref{Figure4}(a) and \ref{Figure4}(b)], we compute the scattering parameters, assuming a normally incident, TM-polarized plane-wave illumination at 14.5 GHz. In our simulations, the standard (MUMPS) solver is utilized, and the structure is discretized via an adaptive mesh with default parameters and maximum  element size of $\lambda_0/100$, which results in about 120,000 degrees of freedom. Here, and henceforth, the inclusions are assumed as perfectly electric conducting. From the computed scattering parameters, the effective constitutive parameters are extracted by following the standard procedure detailed in \cite{Chen:2009de, Smith:2002do}.

%
\begin{figure}
\centering
\includegraphics[width=8.5cm]{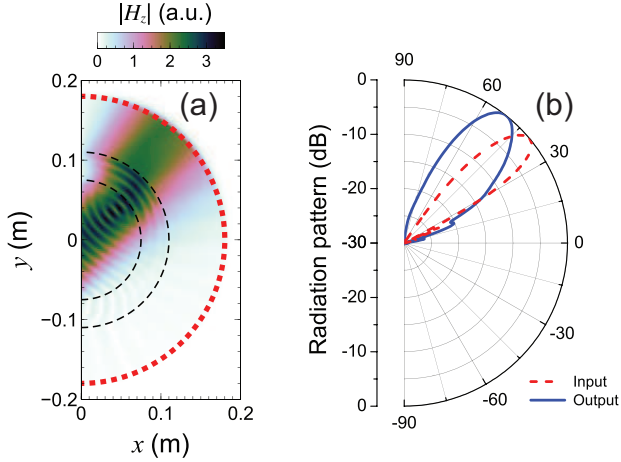}
\caption{(a) Numerically computed near-field map ($|H_z|$ in arbitrary units) at the reference frequency of 14.5 GHz, assuming the ideal TO-based constitutive blueprints in (\ref{eq:TrasnfM}), with $\alpha=1.25$, $R_1=75$ mm, $R_2=110$ mm. The structure is excited by an aperture field distribution of width 148 mm, located at $x=0$, with a linear-phasing and a Taylor-type amplitude taper yielding an incident beam along the direction $\theta_i=40^o$ and a SLL not exceeding -17 dB. The black-dashed contours delimit the radome region, whereas the red-dotted contour delimits the computational domain. Only the relevant ($x>0$) half of the computational domain is shown. (b) Corresponding radiation pattern, with the output beam (blue-solid curve) pointing at $\theta_o=52^o$. Also shown, as a reference, is the input beam (red-dashed curve) pointing at $\theta_i=40^o$. Only the relevant angular range ($-90^o<\theta<90^o$) is shown.}
\label{Figure6}
\end{figure}

%
\begin{figure}
\centering
\includegraphics[width=8.5cm]{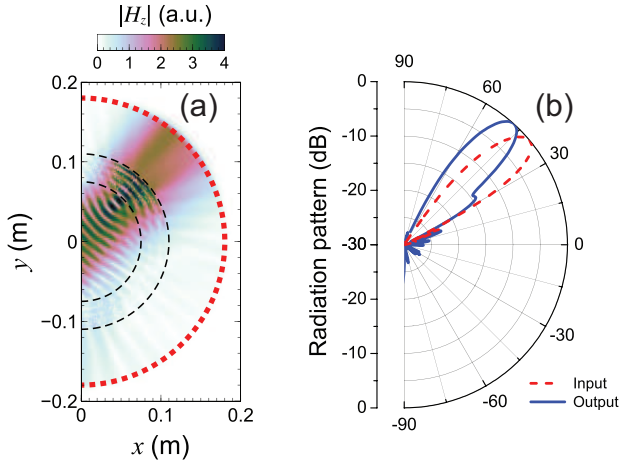}
\caption{As in Fig. \ref{Figure6}, but assuming the metamaterial implementation illustrated in Fig. \ref{Figure4}. 
The output beam in panel (b) is now pointing at $\theta_o=48^o$.}
\label{Figure7}
\end{figure}

For computing the EM response (near-field map and radiation pattern) of the TO-based and synthesized-metamaterial structures, shown hereafter, we consider a circular computational domain of radius 180 mm, terminated by a perfectly matched layer with thickness of 20 mm. The phased-array excitation is simulated via an aperture field distribution   of width 148 mm, located at $x=0$ (see the schematic in Fig. \ref{Figure1}), with a linear-phasing and a Taylor-type amplitude taper \cite{Balanis:2012an} yielding an incident beam along a desired input direction $\theta_i$ and a SLL not exceeding -17 dB. Also in this case, we utilize the standard solver and an adaptive meshing with maximum element size set to $\lambda_0/20$, which yields about 3 million degrees of freedom. From the near-field distribution,
the SWR is estimated as the ratio between the (average) maximum and minimum magnitude values of the standing-wave profile within the inner region $r<R_1$, whereas the radiation pattern is computed via the embedded \texttt{farfield} option in COMSOL Multiphysics.

\subsection{Results}
Figure \ref{Figure6} shows the response of the radome, for an incident beam with $\theta_i=40^o$, when the ideal TO-based constitutive blueprints in (\ref{eq:TrasnfM}) are assumed. From the near-field map [Fig. \ref{Figure6}(a)], we can observe the steering effect and, from the standing-wave pattern in the inner region $r<R_1$ we can estimate a $\mbox{SWR}\approx 1.5$, which is in line with the prediction in (\ref{eq:SWR}). The angular enhancement is more apparent in the far-field pattern [Fig. \ref{Figure6}(b)], with the output beam pointing at $\theta_o=52^o$, in accord with the targeted factor $\alpha=1.25$. 

Figure \ref{Figure7} shows instead the response obtained by assuming the actual metamaterial implementation based on 
thin-metallic-plate inclusions (see Fig. \ref{Figure4}). Results look qualitatively similar to the previous case. 
The standing-wave pattern in the inner region $r<R_1$ [Fig. \ref{Figure7}(a)] is slightly more pronounced ($\mbox{SWR}\approx 2$), which is expected and can be attributed to the relaxed synthesis procedure (see the discussion in Sec. \ref{Sec:MTM-synthesis}). Nevertheless, the radome correctly implements the required steering, with the output beam [Fig. \ref{Figure7}(b)] now directed along $\theta_o=48^o$, once again consistent with the targeted enhancement factor $\alpha=1.25$.
The coma-type aberrations that are observable in the output beams of Figs. \ref{Figure6} and \ref{Figure7} are likely attributable to the moderate size (about seven wavelengths) of the array aperture as well as the distortions induced by the imperfect impedance matching.

%
\begin{figure}
\centering
\includegraphics[width=8.5cm]{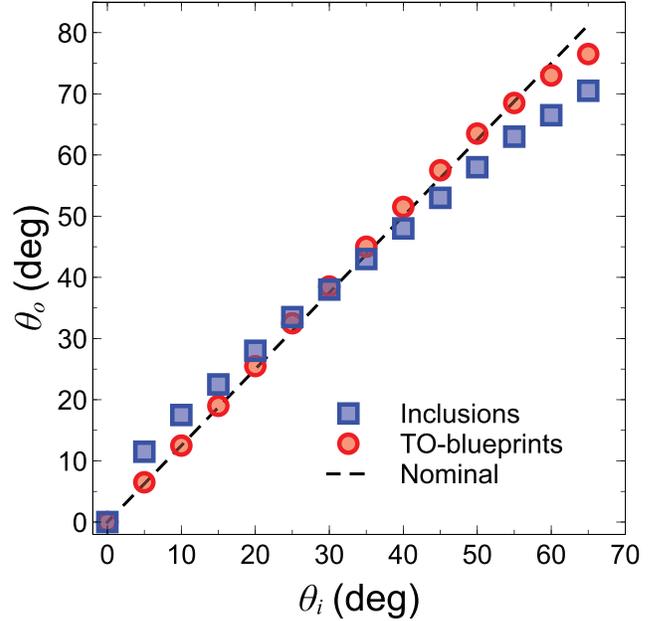}
\caption{Input-output characteristics for the ideal (red-circle markers) and inclusion-based (blue-square markers) radomes at the reference frequency of 14.5 GHz. The black-dashed line indicates the nominal characteristic in (\ref{eq:steering}) (with $\alpha=1.25$).}
\label{Figure8}
\end{figure}

%
\begin{figure}
\centering
\includegraphics[width=8.5cm]{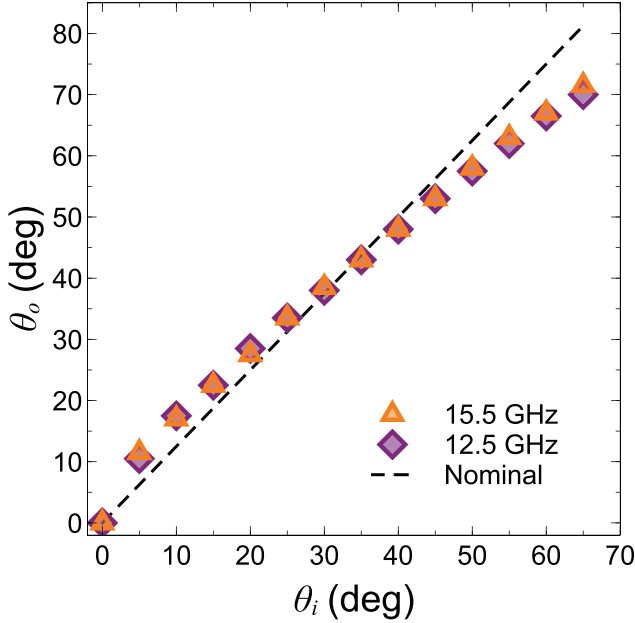}
\caption{As in Fig. \ref{Figure8}, but for the inclusion-based structure operating at 12.5 GHz (purple-diamond markers) and 15.5 GHz (orange-triangle markers).}
\label{Figure9}
\end{figure}

%
\begin{figure}
\centering
\includegraphics[width=8.5cm]{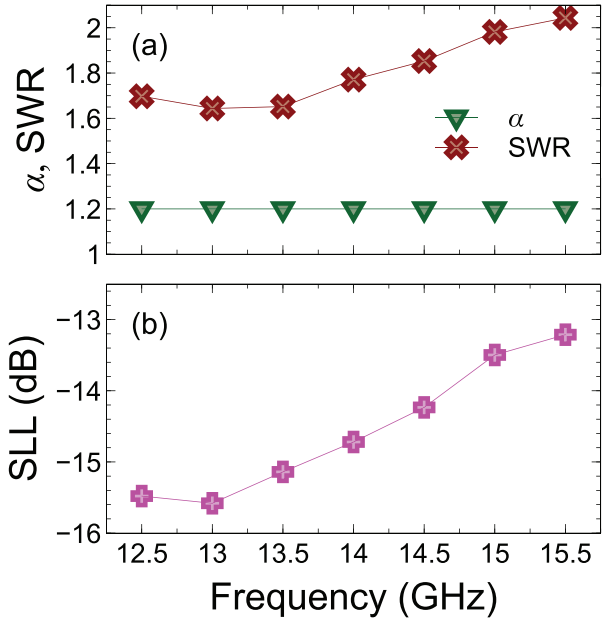}
\caption{(a) Enhancement factor (green-triangle markers) and SWR (dark-red-cross markers) for the inclusion-based structure as a function of frequency, for $\theta_i=40^o$. (b) Corresponding SLL. Continuous curves are guides to the eye only.}
\label{Figure10}
\end{figure}

For the ideal and inclusion-based structures, Fig. \ref{Figure8} summarizes the input-output characteristics ($\theta_o$ as a function of $\theta_i$), within a range of incident directions ($0\le \theta_i\le 65^o$) for which the input beam maintains a sufficiently narrow and well-defined profile. As it can be observed, the results pertaining to the ideal TO-based blueprints closely follow the nominal characteristic. Those pertaining to the synthesized metamaterial exhibit a mild nonlinearity, with a slope that is slightly larger than the nominal one for smaller angles, and moderately decreases for larger angles. These departures are attributable to the various simplifications and approximations introduced in the metamaterial synthesis, and may be overcome by affording more complex implementations, e.g., by utilizing inclusion geometries featuring additional degrees of freedom.   
Nonetheless, it is remarkable that our simple proof-of-principle implementation in Fig. \ref{Figure5} (featuring only two types of inclusions) captures the essential angle-enhancement functionality to a fairly reasonable extent.
We also point out that the observed sidelobe level (SLL) in the output beams of all examples above is $< -15$ dB on average, and never exceeding $-12$ dB.

One attractive feature of our proposed approach and implementation is their potentially broadband character, since they do not rely on resonant constituents. To illustrate this aspect, Fig. \ref{Figure9} shows the input-output characteristics of the inclusion-based structure at 12.5 GHz and 15.5 GHz, which are very similar to those in Fig. \ref{Figure8}. 
Figure \ref{Figure10} shows a representative behavior (for $\theta_i=40^o$) of the enhancement factor, SWR and SLL within the above frequency range. It can be observed that the enhancement factor remains constantly close to the targeted value ($\alpha=1.25$), and the levels of SWR and SLL remain satisfactorily low.
At frequencies higher than 15.5 GHz, the output beam for $\theta_i=0$ exhibits a slight splitting, which is likely attributable to the particularly aggressive decimation of the inclusions in the central region of the radome (see Fig. \ref{Figure5}). At frequencies lower than 12.5 GHz, the input beam ceases to exhibit a well-defined profile for larger angles ($\theta_i\sim 60^o$), and so the corresponding results are not meaningful. Nevertheless, even though no specific optimization was performed in this respect, a bandwidth of over 20\% is attained. 

\section{Conclusions}
\label{Sec:Conclusions}
To sum up, we have presented a TO-based design of a metamaterial radome for extending the scanning angle of phased-array antennas. In particular we have introduced a new coordinate transformation which, via suitable parameter optimization, 
yields some constitutive blueprints that do not involve negative and/or extreme-parameter media, and are compatible with a metamaterial synthesis based on thin-plate-metallic inclusions.
We have characterized the synthesized metamaterial structure via full-wave numerical simulations, illustrating its robustness and bandwidth properties, as well as its current limitations. Overall, the attained synthesis provides a proof-of-principle demonstration that paves the way to a technologically viable, robust, and potentially broadband implementation. Within this framework, the fabrication and experimental characterization of a Ku-band prototype is currently under way.

Among the possible future developments, it is worth mentioning the study of new classes of coordinate transformations that can mitigate the impedance-mismatch. For instance, a first-order impedance-matching condition at the boundary $r=R_2$ could be realized by relaxing the identity condition in (\ref{eq:c2}). Also of interest is the exploration of
different radome shapes and arbitrary polarizations, as well as metallo-dielectric implementations that are better suited to additive fabrication technologies such as 3-D printing. 

\appendix[Impedance Mismatch]
In the auxiliary vacuum domain [Fig. \ref{Figure1}(a)], the radiated magnetic field can be expanded in terms of angular-momentum modes
\beq
H_{zm}^{\prime}\left(r^\prime,\theta^\prime\right)=\mbox{H}_m^{(1)}\left(k_0r^\prime\right)
\exp\left(i m \theta^\prime\right),
\eeq
where $\mbox{H}_m^{(1)}$ denotes the $m$-th order Hankel function of the first kind \cite{Abramowitz:1964}, and $k_0=2\pi/\lambda_0$ the vacuum wavenumber.
In the transformed domain [Fig. \ref{Figure1}(b)], these modes are mapped [via (\ref{eq:CTr})] as
\beq
H_{zm}\left(r,\theta\right)=\mbox{H}_m^{(1)}\left[k_0 F_r\left(r\right)\right]
\exp\left[i m F_{\theta}\left(r\right)\theta\right].
\label{eq:Hzm}
\eeq
The corresponding $\theta$-directed electric fields can be obtained via the relevant Maxwell's curl equation, viz.,
\beq
E_{\theta m}\left(r,\theta\right)=\frac{i\eta_0}{k_0}\left\{
{\underline {\underline \varepsilon}}^{-1}
\left(r,\theta\right)\cdot
\nabla\times\left[H_{zm}\left(r,\theta\right) {\hat {\bf u}_z}\right]\right\}\cdot {\hat {\bf u}_{\theta}}
\label{eq:Ethetam}
\eeq
with ${\underline {\underline \varepsilon}}$ given in (\ref{eq:TrasnfM}), $\eta_0=\sqrt{\mu_0/\varepsilon_0}\approx 377 \Omega$ denoting the vacuum characteristic impedance, and ${\hat {\bf u}_z}$ and ${\hat {\bf u}_{\theta}}$ unit-vectors. From (\ref{eq:Hzm}) and (\ref{eq:Ethetam}), via some algebra, it follows that a modal impedance can be defined as
\beq
Z_m\left(r\right)=\frac{E_{\theta m}\left(r,\theta\right)}{H_{zm}\left(r,\theta\right)}=
 - \frac{i\eta_0 F_r\left(r\right) F_{\theta}\left(r\right)
 \dot{\mbox{H}}_m^{(1)}\left[k_0 F_r\left(r\right)
 \right]}
 {r \mbox{H}_m^{(1)}\left[k_0 F_r\left(r\right)\right]}.
\label{eq:Zm}
\eeq
By exploiting in (\ref{eq:Zm}) the large-argument asymptotic expansion of the Hankel function (and its derivative) \cite{Abramowitz:1964}, we derive the following approximation
\beq
Z_m\left(r\right)
\sim \eta_0\frac{F_r\left(r\right)F_{\theta}\left(r\right)}{r},
\eeq
which, particularized at the two boundaries delimiting the radome region yields [taking into account (\ref{eq:BC})]
\beq
Z_m\left(R_1\right)\sim \eta_0,
\eeq
\beq
Z_m\left(R_2\right)\sim \alpha \eta_0.
\label{eq:Zm2app}
\eeq
Within the validity range of the above approximation, the wave impedance of the angular-momentum modes is independent of the order $m$, and varies within the radome region, starting from a value  that is matched with vacuum (at $r=R_1$), and growing by a factor that is approximately equal to the angular enhancement factor $\alpha$ (at $r=R_2$). It is important to point out that the inherent  $\theta$-dependence of the coordinate transformation in (\ref{eq:CTr}) implies a coupling of the angular-momentum modes, and therefore it is not straightforward to relate the actual wave impedance (which is likewise $\theta$-dependent) to the modal ones. Nevertheless, for moderate values of the angular enhancement factor $\alpha$, for which the modal coupling can be expected to be not particularly strong, the approximation in (\ref{eq:Zm2app}) may be used as a rough estimate of the (average) wave impedance. This yields in turn the SWR estimate in (\ref{eq:SWR}), which has been verified to be in line with the results obtained via full-wave numerical simulations.

\section*{Acknowledgment}
This work has been supported in part by the
Italian Ministry of Education, University and Research (MIUR) through the Campania Aerospace District 
within the framework of the TELEMACO project 
(PON03PE-00112-1) ``Enabling technologies and innovative electronic 
scanning systems in millimeter and centimeter bands for avionic radar applications.''

\bibliographystyle{IEEEtran}
\bibliography{IEEE-JMMCT_MTM-radome}

\end{document}